\documentclass[aps,reprint,10pt,notitlepage,pra,
               onecolumn,
               amsfonts,amssymb,amsmath,
               showpacs,
               nobibnotes,
               nofootinbib,
               ]{revtex4-1}
\usepackage{graphicx}
\usepackage{amssymb}
\usepackage{amstext}
\usepackage{amsmath}
\usepackage{amsthm}
\usepackage[warn]{mathtext}
\usepackage{braket}
\usepackage{bm}
\usepackage{indentfirst}
\usepackage{wrapfig}
\usepackage{topcapt}
\usepackage{comment}
\usepackage{subfigure}
\usepackage{color,xcolor}
\definecolor{dark-blue}{rgb}{0,0,0.875}
\definecolor{dark-green}{rgb}{0,0.625,0}
\definecolor{dark-red}{rgb}{0.875,0,0}
\usepackage[colorlinks=true,linkcolor=dark-red,filecolor=dark-blue,
citecolor=dark-green,urlcolor=dark-blue,breaklinks=true]{hyperref}
\begin{document}
\title{Generation of gaussian entangled states of light in an array of nonlinear waveguides}
\author{V. O. Martynov}
\affiliation{Institute of Applied Physics of RAS, Nizhny Novgorod, Russia}
\author{V. A. Mironov}
\affiliation{Institute of Applied Physics of RAS, Nizhny Novgorod, Russia}
\author{L. A. Smirnov}
\affiliation{Institute of Applied Physics of RAS, Nizhny Novgorod, Russia}
\affiliation{Department of Control Theory, Nizhny Novgorod State University,
Gagarin Av. 23, 606950, Nizhny Novgorod, Russia}
\pacs{42.65.Lm , 42.65.W, 42.82.-m, 03.67.Bg}

\begin{abstract}
We investigate process of entangled state of light generation while propagation along a one dimensional array of single-mode nonlinear waveguides. We consider a situation when entanglement is formed due to spontaneous parametric down-conversion of the pump which is present only in a signal waveguide. In the considered process the generated state of light is multi-mode squeezed. We demonstrate that starting from certain distance of light propagation only pairs of waveguides, located symmetrically with respect to the pumped one, occur to be entangled. Also there is an optimal pump amplitude for which the formed quantum correlations are most pronounced. Entanglement for multi-mode squeezed states may be very sensitive for phase fluctuations in the pump. We investigate the influence of such noise on the discussed process. We demonstrate that for situation of generation of few photon entangled states the influence of phase fluctuations is negligible. But it dramatically increase with the growth of average photon numbers in the formed quantum states.
\end{abstract}
\maketitle
\section{Introduction}
The use of optical waveguides is currently being actively studied in context of various problems of quantum optics.
From applications point of view, of course, the most interesting possibility is the use of such systems for implementing various quantum informatics algorithms~\cite{nielsen_quantum_2000, braunstein_quantum_2003}.
For certain problems, these algorithms allow to get a performance gain compared to classical computing systems.
Despite more than thirty years of research in this area, the question of creating an effective physical base for implementing these algorithms is still open.
In this context, advances of the recent years in the development of photonic integrated circuits technology for controlling quantum states of light~\cite{zoubi_quantum_2017, silverstone_-chip_2014, matthews_manipulation_2009, kruse_dual-path_2015} and their tomography~\cite{solntsev_path-entangled_2017} look promising.
One of the most important resources for quantum information processes is the entangled states of quantum systems.
Thus, the generation of entangled states of light is one of the key problems.
This problem has been successfully solved for many years using crystals with quadratic nonlinearity~\cite{kolobov_quantum_2007, zerom_entangled-photon_2011, ueno_entangled_2012}. 
In spite of this, from a technological point of view, it is important to develop sources of entangled states of light integrated into waveguide circuits.
This problem has been discussed in a number of recent papers~\cite{peruzzo_quantum_2010, meinecke_coherent_2013, guo_parametric_2017, caspani_integrated_2017}. 
For example, in~\cite{solntsev_generation_2014,antonosyan_effect_2014, solntsev_path-entangled_2017,yang_manipulation_2014}, the possibility of entangled states generation as a result of spontaneous parametric down-conversion in a one-dimensional array of coupled waveguides with quadratic nonlinearity is studied theoretically and experimentally.
In particular, the situation when the optical field on pump frequency present only in one wavequide is considered there.
As a result of the pump down conversion, photon pairs on the near half pump frequency appear in the array. 
Quantum walks in the transverse direction of these photon pairs form entanglement between individual waveguides.
A feature of the previously mentioned papers is that the process of generation of only two-photon states is considered there.
From the point of view of quantum informatics, the advantage of such states is that for them it is possible to perform high fidelity quantum operations required for different algorithms.
However, to generate such states in the process of spontaneous parametric down-conversion, it is necessary to work in low pump power regime to reduce the probability of states with a larger number of photons.
This fact strongly limits the operation frequency of quantum devices based on such states. \par
An alternative way to the use of light in the quantum informatics is based on Gaussian states (for example, coherent or squeezed), in which information is encoded in quadratures of electromagnetic fields~\cite{braunstein_quantum_2003, weedbrook_gaussian_2012, masada_continuous-variable_2015}.
In general case in spontaneous parametric down conversion process the quantum state of generated light is squeezed. 
This mean that it is possible to construct deterministic source of entangled states in this, so called continuous variables, approach. 
But still, it should be mentioned that there are questions to possible efficiency of quantum algorithms based on such states~\cite{leverrier_security_2017, ichikawa_notes_2017}.
In this paper, we will discuss the process of the formation of multimode squeezed states in a one-dimensional array of coupled waveguides with quadratic nonlinearity.
Mainly, we will discuss the entanglement formed between different optical modes.
It should be noted here that, as shown in~\cite{martynov_influence_2017}, even small phase fluctuations in the pump can significantly limit the formation of Gaussian entangled states in parametric systems.
Thus, in our work we also consider the influence of such fluctuations on the entanglement evolution in the system.
In Section 2, we describe the theoretical model, we give the equations describing the evolution of the state of light during propagation along the waveguides array, and describe the method of quantifying of entanglement we use and give all the necessary formulas.
In Section 3 we discuss the obtained results.
And finally, in section 4 we summarize.
\section{Theoretical model}
\subsection{Base equations}
Let's consider a one dimensional array of single-mode optical waveguides with quadratic nonlinearity~(see Fig.\ref{plot:schema}).
Phase matching for the process of spontaneous parametric down conversion of a photon with a frequency $\omega_p$ into two photons with a frequencies $\omega_s = \omega_p/2$ is performed in each fiber.
We assume that the spectrum of the field in each waveguide contains only components with frequencies close to $\omega_p$ and $\omega_s$.
Thus, each of these frequencies can be described by a slowly varying operator: $ \hat{a}^{p}_{n}\left(t,z\right)$ - for frequency $\omega_p$, $ \hat{a}^{s}_{n}\left(t,z\right)$ - for frequency $\omega_s$.
Index $n$ defines  optical waveguide in the array.
The array arranged in such a way that modes of neighboring waveguides interacts. 
In the weak coupling approximation, the equations on slowly varying operators have the following form:
\begin{figure}[t]
	\begin{minipage}[t]{1.0\linewidth}
		\includegraphics[scale=1.0]{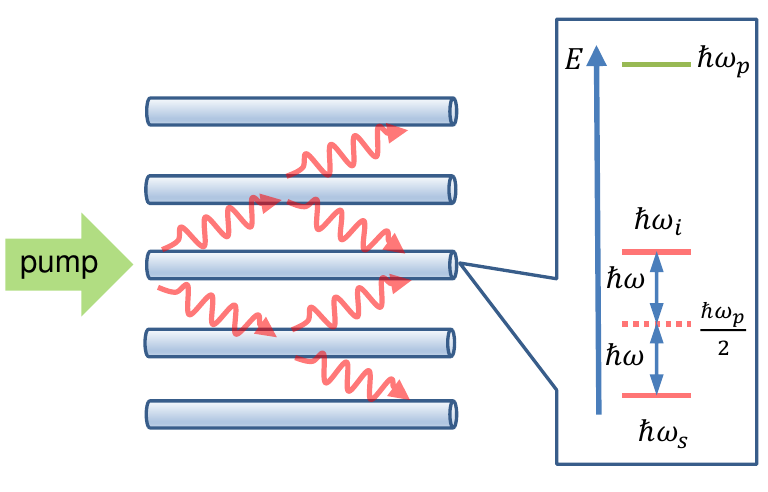}\vspace{-1.25mm}
		\caption{Principal schema of the discussed array of nonlinear waveguides.}
		\label{plot:schema}
	\end{minipage}
\end{figure}
\begin{eqnarray}
\label{equation:pump:1}
\left(\dfrac{\partial}{\partial z} + \dfrac{1}{v_{p}}\dfrac{\partial}{\partial t}\right)\hat{a}^{p}_{n}\left(t,z\right) = iC_{p}\left(\hat{a}^{p}_{n-1}\left(t,z\right) +\hat{a}^{p}_{n+1}\left(t,z\right)\right) + 2i\chi \hat{a}^{s2}_{n},\\
\label{equation:signal:1}
\left(\dfrac{\partial}{\partial z} + \dfrac{1}{v_{s}}\dfrac{\partial}{\partial t}\right)\hat{a}^{s}_{n}\left(t,z\right) = iC_{s}\left(\hat{a}^{s}_{n-1}\left(t,z\right) +\hat{a}^{s}_{n+1}\left(t,z\right)\right) + i\chi \hat{a}^{p}_{n}\hat{a}^{s\dag}_{n}.
\end{eqnarray}
Here $v_p$ and $v_s$ - group velocity for corresponding frequency; $C_{p}$, $C_{s}$ - interaction coefficients of neighbouring waveguides modes, $ \chi$ -  coefficient of quadratic nonlinearity. 
For higher frequency the overlap integral of modes decrease exponentially.
Therefore, inequality $C_{p}\ll C_{s}$ is valid, and we can assume $C_{p}=0$. 
In this paper, we consider the situation, when light at the pump frequency~$\omega_p$ has large intensity on the input of the array.
Moreover, we assume that the length of the array is small enough, so that we can neglect the exhaustion of the pump.
Thereby, operators $\hat{a}^p_n(t,z)$ may be replaced by c-numbers, and for the pump field we get the equation:
\begin{equation}\label{equation:pump:2}
\left(\dfrac{\partial}{\partial z} + \dfrac{1}{v_{p}}\dfrac{\partial}{\partial t}\right)a^{p}_{n}\left(t,z\right) = 0.
\end{equation}
We assume that continuous laser radiation is used as pumping, and it is applied only to one waveguide, which we will refer as central, i.{\,}e $ a^{p}_{n}\left(t,z\right) = 0$ if $ n \ne 0$.
Any source of such radiation has a finite spectral width.
To describe such a spectrum, one can use the phase diffusion model~\cite{scully_quantum_1997}.
According to this model, the field at the input of the array has the next form:
\begin{equation}
a^{p}_{0}\left(t,0\right) = A\cdot e^{i\varphi\left(t\right)},
\end{equation} 
where $A$ - pump amplitude, and $\varphi\left(t\right)$ - random Wiener process with zero mean value and the following correlation relations: 
\begin{equation}
\langle \dot{\varphi}\left(t\right)\dot{\varphi}\left(t'\right)\rangle_{\varphi} = 2\cdot\Delta\omega\cdot\delta\left(t-t'\right),
\end{equation}
where a dot over a function means the time derivative; averaging is performed over an ensemble of realizations of $\varphi$; $\Delta\omega$ - spectral width of the pump. Solution of equation (\ref{equation:pump:2}) for central waveguide has the next form:
\begin{equation}
a^{p}_{0}\left(t,z\right) = A\cdot e^{i\varphi\left(t - z/v_{p}\right)}.
\end{equation}
Substitute this solution into equation (\ref{equation:signal:1}) and this lead to:
\begin{equation}\label{equation:signal:2}
\left(\dfrac{\partial}{\partial z} + \dfrac{1}{v_{s}}\dfrac{\partial}{\partial t}\right)\hat{a}^{s}_{n}\left(t,z\right) = iC_{s}\left(\hat{a}^{s}_{n-1}\left(t,z\right) +\hat{a}^{s}_{n+1}\left(t,z\right)\right) + ig\delta_{0n}e^{i\varphi\left(t - z/v_{p}\right)}\hat{a}^{s\dag}_{n},
\end{equation} 
where $g = A\cdot \chi$, $\delta$ - Kronecker delta. Next, we perform Fourier transform of equation (\ref{equation:signal:2}):
\begin{equation}\label{equation:signal:3}
\dfrac{\partial}{\partial z}\hat{a}^{s}_{n}\left(\omega,z\right) + \dfrac{i\omega}{v_{s}}\hat{a}^{s}_{n}\left(\omega,z\right) = iC_{s}\left(\hat{a}^{s}_{n-1}\left(\omega,z\right) +\hat{a}^{s}_{n+1}\left(\omega,z\right)\right) + \dfrac{ig\delta_{0n}}{2\pi}\int \hat{a}_{n}^{s\dag}\left(\omega',z\right)U\left(\omega' + \omega,z\right)d\omega',
\end{equation}
where
\begin{eqnarray}
\label{fourier:operator}
\hat{a}^{s}_{n}\left(\omega,z\right) = \int \hat{a}^{s}_n\left(t,z\right)e^{-i\omega t}dt,\\
\label{fourier:function}
U\left(\omega,z\right) = \int e^{i\varphi\left(t - z/v_{p}\right) - i\omega t}dt.
\end{eqnarray}
To simplify expression (\ref{equation:signal:3}) we should note that frequency scale of operators $\hat{a}^{s}_{n}(\omega,z)$ is determined by parametric resonance width, and frequency scale of function $ U(\omega,z)$ is determined by spectral width of the pump. 
We suppose that the spectral width of the pump is much smaller than the width of parametric resonance.
In this case integral in (\ref{equation:signal:3}) equal to:
\begin{equation}
\dfrac{1}{2\pi}\int \hat{a}_{n}^{s\dag}\left(\omega',z\right)U\left(\omega' + \omega,z\right)d\omega' \approx \dfrac{\hat{a}_n^{s\dag}\left(-\omega,z\right)}{2\pi}\int U\left(\omega',z\right)d\omega',
\end{equation} 
thus we get the following equation:
\begin{equation}\label{equation:signal:4}
\dfrac{\partial}{\partial z}\hat{a}^{s}_{n}\left(\omega,z\right) + \dfrac{i\omega}{v_{s}}\hat{a}^{s}_{n}\left(\omega,z\right) = iC_{s}\left(\hat{a}^{s}_{n-1}\left(\omega,z\right) +\hat{a}^{s}_{n+1}\left(\omega,z\right)\right) + ig\delta_{0n}e^{i\varphi\left(- z/v_{p}\right)}\hat{a}^{s\dag}_{n}\left(-\omega,z\right).
\end{equation}
For a full description of the evolution of light, equations~(\ref{equation:signal:4}) should be supplemented by quantum state of light on the input of the array.
Throughout this paper we assume that for frequencies, close to $\omega_s$, quantum state is vacuum 
Thus, we have described a formalism that allows us to fully determine the state of light in an arbitrary place of the array.
In particular, we can calculate the averages of any combinations of creation and annihilation operators, as well as determine the amount of entanglement in the waveguide system. This will be discussed in the next section.
\subsection{Calculation of entanglement evolution}
The discussed system consists of a large number of quantum optical modes.
For such a system, the complete determination of all present quantum correlations is quite a challenge.
In our work, we use a simple method, which consists in calculating the amount of entanglement only for a pair of modes, while quantum mechanical averaging over the rest of the system is performed.
Going through all possible pairs, we get a distribution, which we will analyze.
This method is often used to study complex quantum systems consisting of a large number of particles~\cite{solntsev_generation_2014, tura_detecting_2014, mazza_detecting_2015, laflorencie_quantum_2016}. 
However, it is possible to miss a certain kind of quantum correlation while using such method. For example if the quantum state of the system is a GHZ-like \par
To determine entanglement between two optical modes, any two-particle criterion can be used.
Since all of them are equivalent, the results qualitatively will be the same.
Hereinafter, we will use a quantity called logarithmic negativity $E_N$~\cite{vidal_computable_2002}.
On the input of the array we use vacuum quantum state for modes having frequency near $\omega_s$.
In addition, due to linearity of equations~(\ref{equation:signal:4}) quantum state of light will be Gaussian in any place of the array.
For this case, there is a simplified algorithm for calculating logarithmic negativity.
Let's consider a system consisting of two optical modes.
The first mode is described by the creation (annihilation) operators $\hat{a}^{\dag}$ ($\hat{a}$), and the second one by $\hat{b}^{\dag}$ ($\hat{b}$).
Build a matrix:
\begin{eqnarray}\label{eq:CovarianceMatrix1}
\sigma_{m,n}=\langle\hat{\xi}_{m}\hat{\xi}_{n}+\hat{\xi}_{m}\hat{\xi}_{n}\rangle\bigl/2\bigr.-\langle\hat{\xi}_{m}\rangle\langle\hat{\xi}_{n}\rangle,
\end{eqnarray}
where indexes $m$ and $n$ take value from $1$ to $4$, and $\hat{\xi}_{m}$ and $\hat{\xi}_{n}$ are the corresponding components of a four-dimensional vector $\hat{\bm{\xi}}=\left(\,\hat{q}^{a},\hat{p}^{a},\,\hat{q}^{b},\,\hat{p}^{b}\,\right)^{T}$. 
Components of the mentioned vector are determined by the dimensionless operators of canonical coordinates $\hat{q}^{a}$, $\hat{q}^{b}$ and $\hat{p}^{a}$, $\hat{p}^{b}$ of two optical modes. 
These quantities are related with the annihilation and creations operators as follows:
\begin{equation}\label{eq:qp}
\hat{q}^{a}=\frac{1}{\sqrt{2}}\left(\hat{a}+\hat{a}^{\dagger}\right),\hspace{2.5mm}\hat{p}^{a}=\frac{i}{\sqrt{2}}\left(\hat{a}^{\dagger}-\hat{a}\right),
\end{equation}
and similarly for the second mode.
From the expression~(\ref{eq:CovarianceMatrix1}) follows directly that $\bm{\sigma}$ can be presented in block representation:
\begin{equation}\label{eq:CovarianceMatrix2}
\bm{\sigma}=
\begin{pmatrix}
\bm{\alpha} && \bm{\gamma} \\
\bm{\gamma}^{T} && \bm{\beta}
\end{pmatrix},
\end{equation}
in which each block $\bm{\alpha}$, $\bm{\beta}$, $\bm{\gamma}$ and $\bm{\gamma}^{T}$ has size $2\times2$. 
The next step is to calculate the values:
\begin{equation}\label{eq:SymplecticInvariants}
A=\det{\bm{\alpha}},\hspace{2.5mm}B=\det{\bm{\beta}},\hspace{2.5mm}C=\det{\bm{\gamma}},\hspace{2.5mm}\varSigma=\det{\bm{\sigma}},
\end{equation}
through which one can express the desired logarithmic negativity:
\begin{equation}\label{eq:LogarithmicNegativity2}
E_{N}=\max\left(0,-\log_{2}\bigl[2\tilde{\nu}_{-}\bigr]\right),
\end{equation}
where 
\begin{equation}\label{eq:nu_minus}
\tilde{\nu}_{-}=\sqrt{\frac{1}{2}\left(\left(A+B-2C\right)-\sqrt{\left(A+B-2C\right)^2-4\varSigma}\right)}.
\end{equation}
\par
Thus, for any pair of optical waveguides it is necessary to calculate all elements of the matrix $\bm{\sigma}$ for given distance from the input of the array. 
After that, using expressions~(\ref{eq:LogarithmicNegativity2}) and~(\ref{eq:nu_minus}), one can calculate logarithmic negativity $E_{N}$.
After that it is possible to build the distribution we have described in the beginning of the current subsection and investigate evolution of entanglement in the system. 
According to~(\ref{eq:CovarianceMatrix1}) and~(\ref{eq:qp}), in general case, $\bm{\sigma}$, corresponding to different pairs, contains different combinations of averaged quadratic combination   
$\langle\hat{a}^{\phantom{\dagger}}_{m}\left(\omega,z\right)\hat{a}^{\phantom{\dagger}}_{n}\left(\omega,z\right) \rangle$, $\langle\hat{a}^{{\dagger}}_{m}\left(\omega,z\right)\hat{a}^{\phantom{\dagger}}_{n}\left(\omega,z\right) \rangle$, $\langle\hat{a}^{\phantom{\dagger}}_{m}\left(\omega,z\right)\hat{a}^{\phantom{\dagger}}_{n}\left(-\omega,z\right) \rangle$ and $\langle\hat{a}^{{\dagger}}_{m}\left(\omega,z\right)\hat{a}^{\phantom{\dagger}}_{n}\left(-\omega,z\right) \rangle$ consisting of annihilation $\hat{a}^{\phantom{\dagger}}_{n}\left(\omega,z\right)$ and creation $\hat{a}^{\dagger}_{n}\left(\omega,z\right)$ operators. Here, as before, indexes $m$ and $n$ point to certain waveguide. 
It is also worth noting that the matrices under consideration also include the averages of the creation (annihilation) operators themselves.
But after averaging~(\ref{equation:signal:4}), the result equations will be a set of homogeneous ordinary differential equations.
The initial conditions for the given equations are zeros, because we use vacuum quantum state as initial.
Thus, the average values of the creation (annihilation) operators are zero for the entire length of the optical waveguides array.\par
To calculate the averages mentioned above, taking into account the presence of phase noise, we use the approach described in~\cite{martynov_influence_2017}.
First of all we should note that in equations~(\ref{equation:signal:4}) $\omega$ is a parameter which doesn't influence evolution of quantum correlations. 
This is because any dependency on this parameter can be removed by simple substitution $\hat{a}^{s}_{n}\rightarrow\hat{a}^{s}_{n}e^{-i\omega/v_{s}}$. 
But still there is a special case $\omega=0$. 
For this degenerate case only single frequency mode participate in the parametric process. 
While in general case there is interaction between two modes in each waveguide which may be referred as signal and idler.  
This circumstance leads to the fact that these two cases are described by different sets of equations for averaged values.
Here we will give equations for both cases.
For the general case, from~(\ref{equation:signal:4}) one can obtain the complete system of equations for the following sets of values:
\begin{equation}
\begin{array}{c}
\hat{\bm{u}}_{m,n}^{(1)}=
\left(\hat{b}^{{\dagger}}_{m}\hat{b}^{\phantom{\dagger}}_{n}, \hat{c}^{{\dagger}}_{m}\hat{c}^{\phantom{\dagger}}_{n} , \hat{b}^{\phantom{\dagger}}_{m}\hat{c}^{\phantom{\dagger}}_{n}e^{-i\varphi}\right),\\
\hat{\bm{u}}_{m,n}^{(2)}=
\left(\hat{b}^{\phantom{\dagger}}_{m}\hat{c}^{\phantom{\dagger}}_{n}, \hat{b}^{{\dagger}}_{m}\hat{b}^{\phantom{\dagger}}_{n}e^{i\varphi} , \hat{c}^{{\dagger}}_{m}\hat{c}^{\phantom{\dagger}}_{n}e^{i\varphi} , \hat{b}^{{\dagger}}_{m}\hat{c}^{{\dagger}}_{n}e^{2i\varphi}\right),\\
\hat{\bm{u}}_{m,n}^{(3)}=
\left(\hat{b}^{{\dagger}}_{m}\hat{c}^{\phantom{\dagger}}_{n},
\hat{c}^{\phantom{\dagger}}_{m}\hat{c}^{\phantom{\dagger}}_{n}e^{-i\varphi},
\hat{b}^{{\dagger}}_{m}\hat{b}^{{\dagger}}_{n}e^{i\varphi}\right),\\
\hat{\bm{u}}_{m,n}^{(4)}=
\left(\hat{b}^{\phantom{\dagger}}_{m}\hat{b}^{\phantom{\dagger}}_{n},
\hat{c}^{{\dagger}}_{m}\hat{b}^{\phantom{\dagger}}_{n}e^{i\varphi},
\hat{b}^{{\dagger}}_{m}\hat{b}^{{\dagger}}_{n}e^{2i\varphi}\right),
\end{array}
\end{equation}
where for brevity we have denoted $\hat{b}_n = \hat{a}^{\phantom{\dagger}}_{n}\left(\omega,z\right)e^{-i\omega/v_{s}}$ and $\hat{c}_n=\hat{a}^{\phantom{\dagger}}_{n}\left(-\omega,z\right)e^{i\omega/v_{s}}$. 
After performing an averaging procedure similar to~\cite{martynov_influence_2017}, as a result, for $\hat{\bm{u}}_{m,n}^{(1)}$ and $\hat{\bm{u}}_{m,n}^{(2)}$ we will obtain:
\begin{equation}\label{eq:non_degenerate:1}
\begin{array}{c}
\left(u^{(1)}_{1,m,n}\right)^{'}_z\!=\!
iC_s\!\left(u^{(1)}_{1,m,n-1}\!+\!u^{(1)}_{1,m,n+1}\!-\!u^{(1)}_{1,m-1,n}\!-\!u^{(1)}_{1,m+1,n}\right)\! + \\
+ ig\delta_{0n} u^{(1)*}_{3,m,n}\!-\!ig\delta_{0m}u^{(1)}_{3,n,m}\, ,\\

\left(u^{(1)}_{2,m,n}\right)^{'}_z\!=\!
iC_s\!\left(u^{(1)}_{2,m,n-1}\!+\!u^{(1)}_{2,m,n+1}\!-\!u^{(1)}_{2,m-1,n}\!-\!u^{(1)}_{2,m+1,n}\right)\! + \\
+ ig\delta_{0n} u^{(1)*}_{3,n,m}\!-\!ig\delta_{0m}u^{(1)}_{3,m,n}\, ,\\

\left(u^{(1)}_{3,m,n}\right)^{'}_z\!=\! -\dfrac{\Delta\omega}{v_p}u^{(1)}_{3,m,n}\! +\!
iC_s\!\left(u^{(1)}_{3,m,n-1}\!+\!u^{(1)}_{3,m,n+1}\!+\!u^{(1)}_{3,m-1,n}\!+\!u^{(1)}_{3,m+1,n}\right) \!+ \\
+ ig\delta_{0n} u^{(1)}_{1,n,m}\!+\!ig\delta_{0m}u^{(1)}_{2,m,n}\! +\! ig\delta_{0m}\delta_{0n}\, ,\\
\end{array}
\end{equation}
\begin{equation}\label{eq:non_degenerate:2}
\begin{array}{c}
\left(u^{(2)}_{1,m,n}\right)^{'}_z\!=\! 
iC_s\!\left(u^{(2)}_{1,m,n-1}\!+\!u^{(2)}_{1,m,n+1}\!+\!u^{(2)}_{1,m-1,n}\!+\!u^{(2)}_{1,m+1,n}\right)\! + \\
+ ig\delta_{0n} u^{(2)}_{2,n,m}\!+\!ig\delta_{0m}u^{(2)}_{3,m,n}\! +\! ig\delta_{0m}\delta_{0n}e^{-\dfrac{\Delta\omega}{v_p}z}\, ,\\

\left(u^{(2)}_{2,m,n}\right)^{'}_z\!=\! -\dfrac{\Delta\omega}{v_p}u^{(2)}_{2,m,n}\! +\!
iC_s\!\left(u^{(2)}_{2,m,n-1}\!+\!u^{(2)}_{2,m,n+1}\!-\!u^{(2)}_{2,m-1,n}\!-\!u^{(2)}_{2,m+1,n}\right)\! + \\
+ ig\delta_{0n} u^{(2)}_{4,m,n}\!-\!ig\delta_{0m}u^{(2)}_{1,n,m}\, ,\\

\left(u^{(2)}_{3,m,n}\right)^{'}_z\!=\! -\dfrac{\Delta\omega}{v_p}u^{(2)}_{3,m,n}\! +\!
iC_s\!\left(u^{(2)}_{3,m,n-1}\!+\!u^{(2)}_{3,m,n+1}\!-\!u^{(2)}_{3,m-1,n}\!-\!u^{(2)}_{3,m+1,n}\right)\! + \\
+ ig\delta_{0n} u^{(2)}_{4,n,m}\!-\!ig\delta_{0m}u^{(2)}_{1,m,n}\, ,\\

\left(u^{(2)}_{4,m,n}\right)^{'}_z\!=\! -\dfrac{4\Delta\omega}{v_p}u^{(2)}_{4,m,n}\! -\!
iC_s\!\left(u^{(2)}_{4,m,n-1}\!+\!u^{(2)}_{4,m,n+1}\!+\!u^{(2)}_{4,m-1,n}\!+\!u^{(2)}_{4,m+1,n}\right)\! - \\
- ig\delta_{0n} u^{(2)}_{2,m,n}\!-\!ig\delta_{0m}u^{(2)}_{3,n,m}\! -\! ig\delta_{0m}\delta_{0n}e^{-\dfrac{\Delta\omega}{v_p}z}\, ,
\end{array}
\end{equation}
here $u^{(k)}_{j,m,n}=\langle\hat{u}^{(k)}_{j,m,n}\rangle$, and $\langle\hat{u}^{(k)}_{j,m,n}\rangle$ - jth element of the set $\hat{\bm{u}}^{(k)}_{m,n}$; $*$ - mean complex conjugate. 
For quantities $\hat{\bm{u}}_{m,n}^{(3)}$ and $\hat{\bm{u}}_{m,n}^{(4)}$ result of averaging procedure occur to be homogeneous equations. 
Taking into account selected initial conditions, these equations has zero solution. 
That's why we don't give them here.\par
In degenerate case $\omega=0$ to calculate all required elements for matrix~(\ref{eq:CovarianceMatrix1}) it is enough to derive equations for the next sets of quantities:
\begin{equation}
\begin{array}{c}
\hat{\bm{q}}_{m,n}^{(1)}=
\left(\hat{b}^{{\dagger}}_{m}\hat{b}^{\phantom{\dagger}}_{n},
\hat{b}^{\phantom{\dagger}}_{m}\hat{b}^{\phantom{\dagger}}_{n}e^{-i\varphi}\right),\\
\hat{\bm{q}}_{m,n}^{(2)}=
\left(\hat{b}^{\phantom{\dagger}}_{m}\hat{b}^{\phantom{\dagger}}_{n},
\hat{b}^{{\dagger}}_{m}\hat{b}^{\phantom{\dagger}}_{n}e^{i\varphi},
\hat{b}^{{\dagger}}_{m}\hat{b}^{{\dagger}}_{n}e^{2i\varphi}\right),
\end{array}
\end{equation}
where we have denoted $\hat{b}_n = \hat{a}^{\phantom{\dagger}}_{n}\left(0,z\right)$. 
After averaging procedure we will acquire:
\begin{equation}\label{eq:degenerate:1}
\begin{array}{c}
\left(q^{(1)}_{1,m,n}\right)^{'}_z\!=\! 
iC_s\!\left(q^{(1)}_{1,m,n-1}\!+\!q^{(1)}_{1,m,n+1}\!-\!q^{(1)}_{1,m-1,n}\!-\!q^{(1)}_{1,m+1,n}\right)\! + \\
+ ig\delta_{0n} q^{(1)*}_{2,m,n}\!-\!ig\delta_{0m}q^{(1)}_{2,n,m}\, ,\\

\left(q^{(1)}_{2,m,n}\right)^{'}_z\!=\! -\dfrac{\Delta\omega}{v_p}q^{(1)}_{2,m,n}\! -\!
iC_s\!\left(q^{(1)}_{2,m,n-1}\!+\!q^{(1)}_{2,m,n+1}\!+\!q^{(1)}_{2,m-1,n}\!+\!q^{(1)}_{2,m+1,n}\right)\! + \\
+ ig\delta_{0n} q^{(1)}_{1,n,m}\!+\!ig\delta_{0m}q^{(1)}_{2,m,n}\! +\! ig\delta_{0m}\delta_{0n}\, ,
\end{array}
\end{equation}
\begin{equation}\label{eq:degenerate:2}
\begin{array}{c}
\left(q^{(2)}_{1,m,n}\right)^{'}_z\!=\! 
iC_s\!\left(q^{(2)}_{1,m,n-1}\!+\!q^{(2)}_{1,m,n+1}\!+\!q^{(2)}_{1,m-1,n}\!+\!q^{(2)}_{1,m+1,n}\right)\! + \\
+ ig\delta_{0n} q^{(2)}_{2,n,m}\!+\!ig\delta_{0m}q^{(2)}_{2,m,n}\! +\! ig\delta_{0m}\delta_{0n}e^{-\dfrac{\Delta\omega}{v_p}z}\, ,\\

\left(q^{(2)}_{2,m,n}\right)^{'}_z\!=\! -\dfrac{\Delta\omega}{v_p}q^{(2)}_{2,m,n}\! +\!
iC_s\!\left(q^{(2)}_{2,m,n-1}\!+\!q^{(2)}_{2,m,n+1}\!-\!q^{(2)}_{2,m-1,n}\!-\!q^{(2)}_{2,m+1,n}\right)\! + \\
+ ig\delta_{0n} q^{(2)}_{3,m,n}\!-\!ig\delta_{0m}q^{(2)}_{1,n,m}\, ,\\

\left(q^{(2)}_{3,m,n}\right)^{'}_z\!=\! -\dfrac{4\Delta\omega}{v_p}q^{(2)}_{3,m,n}\! -\!
iC_s\!\left(q^{(2)}_{3,m,n-1}\!+\!q^{(2)}_{3,m,n+1}\!+\!q^{(2)}_{3,m-1,n}\!+\!q^{(2)}_{3,m+1,n}\right)\! - \\
- ig\delta_{0n} q^{(2)}_{2,m,n}\!-\!ig\delta_{0m}q^{(2)}_{2,n,m}\! -\! ig\delta_{0m}\delta_{0n}e^{-\dfrac{\Delta\omega}{v_p}z}\, ,
\end{array}
\end{equation}
The solution of the systems of ordinary differential equations obtained in this section was performed by the 4th order Runge-Kutta method~\cite{garcia_numerical_2000}.
For calculations an array consisting of 512 waveguies was chosen. 
Such choice of the number was done to neglect the influence of boundaries in transverse direction.
\section{Results}
\subsection{Intensity evolution}
\begin{figure}[t]
	\begin{minipage}[t]{1.0\linewidth}
		\includegraphics[scale=1.0]{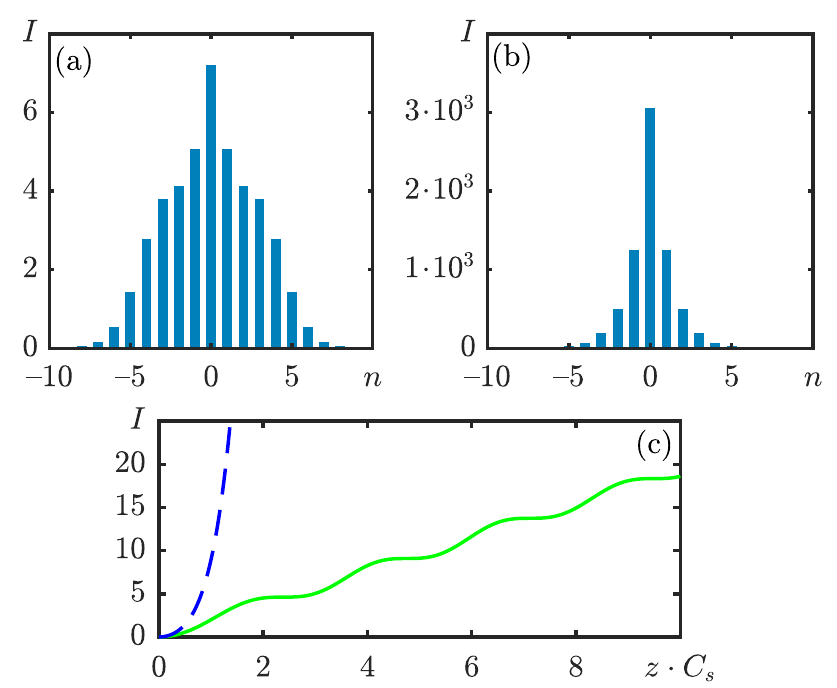}\vspace{-1.25mm}
		\caption{On (a) and (b) distribution of average photon number in waveguide array $I$ is presented for degenerate case when $\omega=0$ in equations~(\ref{equation:signal:4}), $n$ - waveguide index.
		For both figures distribution is calculated for distance $z=3.75/C_{s}$ from the input of the array and pump amplitude corresponds to $g=1.5\cdot C_{s}$ for (a) and $g=2.2\cdot C_{s}$ for (b). 
		On (c) dependency of average photon number in central waveguide on the distance from the input is presented.
		For green solid line $g=1.5\cdot C_{s}$, and for blue dashed $g=2.2\cdot C_{s}$.}
		\label{plot:intensity}
	\end{minipage}
\end{figure}

This section presents the results of the study.
Before we analyze entanglement in the discussed system, we will demonstrate in this subsection some features of light field evolution along the array.
For this we consider change in the average number of photons
$I\left(n,z\right)=\langle\hat{a}_{n}^{\dag}\left(z\right)\hat{a}_{n}\left(z\right)\rangle$ when light propagates along an array of waveguides. 
On Fig. \ref{plot:intensity}(a) and \ref{plot:intensity}(b) distributions calculated based on equations~(\ref{eq:degenerate:1}) are presented. 
Calculations are done for the given distance from array input but different pump amplitude values: $g=1.5\cdot C_{s}$ for (a) and $g=2.2\cdot C_{s}$. 
Also it should be noted that qualitative results we are talking about in this subsection does not depend on either we consider general case or degenerate one. 
So for clarity we have performed calculations for degenerate case.
One may notice that the presented distributions are qualitatively different from each other.
On~\ref{plot:intensity}(b) the average number of photons is much more localized in the vicinity of the central waveguide~(the one which is pumped) as compared to~\ref{plot:intensity}(a).
And this difference becomes more pronounced as the interaction length increases.
This fact indicates a qualitatively different light evolution in the discussed system for the selected two values of the pump amplitude.
This difference is better illustrated in Fig.~\ref{plot:intensity}(c). 
On this figure the dependency of the average number of photons in central waveguide on the propagation distance is presented.
After a more detailed study, it turns out that a qualitative change in light evolution occurs when the pump amplitude exceeds the value~$g=2\cdot C$.
When this occurs, the linear in average increase in the photons number is changed by a rapid exponential growth.
This situation is similar to what is happening in usual quantum damped parametric oscillator. 
It occurs that the system of interacting waveguides we are discussing can be reduced to a parametric oscillator interacting with non-Markovian reservoir~(see Appendix for details).
Unfortunately, the kernel function of result relaxation operator contains Bessel functions.
So we were not able to find any reasonable assumptions to calculate the threshold analytically.
\subsection{Entanglement evolution: coherent pump}
\begin{figure}[t]
	\begin{minipage}[t]{1.0\linewidth}
		\includegraphics[scale=1.0]{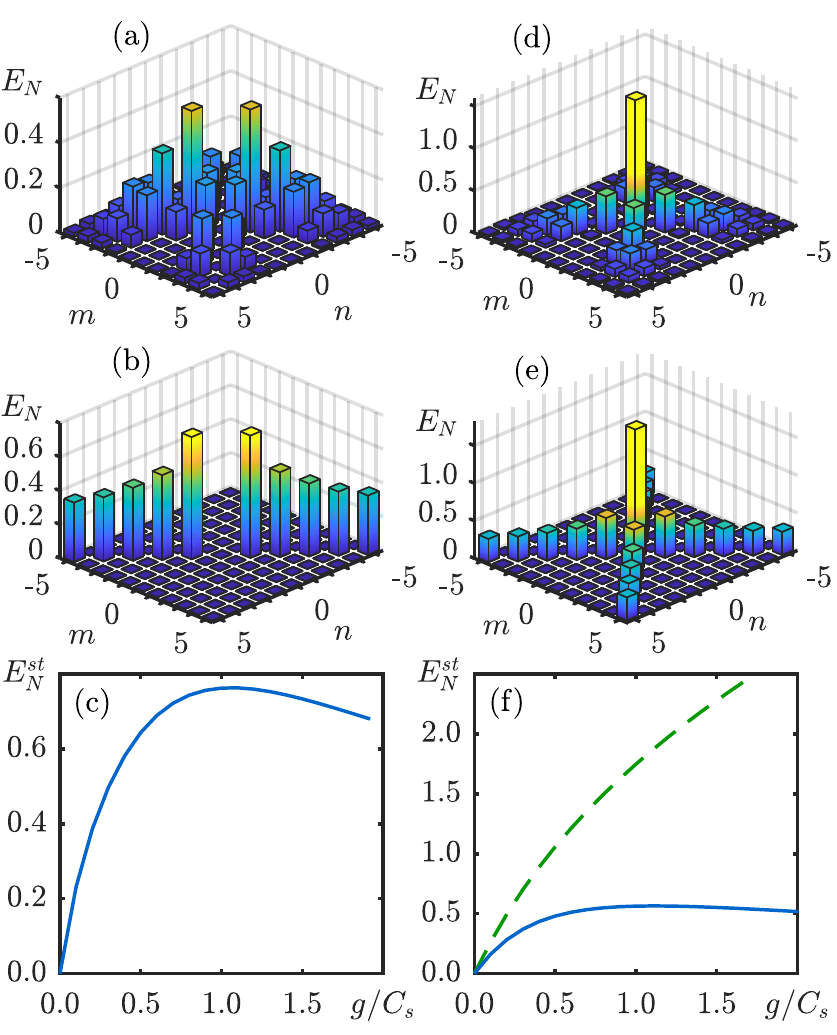}\vspace{-1.25mm}
		\caption{On (a), (b), (d), (e) distributions of logarithmic negativity~$E_N$, calculated for waveguide pairs with indexes n and m, are presented. 
		(a) and (b) correspond to degenerate case for which~$\omega=0$ in~(\ref{equation:signal:4}); (d) and (e) correspond to general case, but for $\omega$ much smaller than parametric resonance width.
		Distributions (d) and (e) are the result of logarithmic negativity calculation for mode pairs with different frequency ( $\omega$ and $-\omega$).
		Each row correspond to a certain distance from the array input: $z=2.25/C_{s}$ for (a) and (d), $z=7.5/C_{s}$ for (b) and (e).
		For all distributions $g=C_{s}$.
		Stationary value of logarithmic negativity dependencies on pump amplitude $g$ are presented on (c) and (f). 
		Figure (c) correspond to degenerate case, and (f) to general.
		In both figures, the solid blue curve corresponds to logarithmic negativity calculated for waveguides with indexes $1$ and $-1$.
		The green dashed line on (f) corresponds to entanglement in the central waveguide between different parametrically coupled spectral components.}
		\label{plot:coherent}
	\end{minipage}
\end{figure}

This subsection is devoted to evolution of entanglement in the array of waveguides for the case of coherent pump, i.{\,}e $\Delta\omega=0$.
Here we use the result of numeric solution of equations ~(\ref{eq:non_degenerate:1}),~(\ref{eq:non_degenerate:2}),~(\ref{eq:degenerate:1}),~(\ref{eq:degenerate:2}) to calculate logarithmic negativity based on~(\ref{eq:LogarithmicNegativity2}),~(\ref{eq:nu_minus}) for different pairs of waveguides.
As a result we obtain distributions, the examples of which are presented on Fig.(a,b,d,e). 
The presented distributions are calculated for the same pump amplitude, but for different distances from the input of the array.
Fig.~\ref{plot:coherent}(a,b) correspond to degenerate case~($\omega = 0$), \ref{plot:coherent}(d,e) correspond to general.
As can be seen from the figures, at the initial stage of entanglement formation quantum correlations exist between a large number of light guide~(Fig.~\ref{plot:coherent}(a,d)).
But as distance increases, the logarithmic negativity becomes zero for all pairs except those located symmetrically relative to the central waveguide~(Fig.~\ref{plot:coherent}(b,e)).
However, in this case, for pairs more distant from the central waveguide, quantum correlations become less pronounced (the value of logarithmic negativity decreases).
This fact is a manifestation of the symmetry of the discussed problem.
A similar phenomenon occurs during the entangled states formation in optical frequency combs~\cite{kues_-chip_2017, roztocki_practical_2017,reimer_generation_2016}.
In such process also entanglement occur to form between the frequency components that are located symmetrically with respect to the pump frequency.
But the difference with our work is that in the system we are investigating entanglement occur to be between spatially separated light modes, and not frequency ones  
Also it should be remarked that distribution in Fig.~\ref{plot:coherent}(d) is similar to the one presented in paper~\cite{solntsev_generation_2014}, where the similar system was discussed but in assumption of fixed total number of photons present in the array. 
Two, to be exact.
This approximation is valid only for arrays with small length, and for a small pump amplitude. \par
Distributions presented on Fig.~\ref{plot:coherent}(b,e) are stationary.
This mean they do not change while further propagation of light along the array and that they  demonstrate the maximum amount of entanglement which can be achieved in the system.
To demonstrate the influence of pump amplitude on formed quantum correlations, dependencies of stationary value of logarithmic negativity calculated for waveguides with indexes $1$ and $-1$ on $g$ are presented on Fig.~\ref{plot:coherent}(c,f). 
It can be seen from the mentioned figures that there is an optimal relationship between pump amplitude and the interaction coefficient of the waveguides $g = 1.1 \cdot C_{s} $.
When this relationship is fulfilled, quantum correlations become most pronounced.
It should be noted that the dependencies are presented only for the values of the pump amplitude not exceeding the threshold $ g = 2 \cdot C_{s} $.
Our calculations showed that stationary value of logarithmic negativity continue decreasing even for pump amplitude above threshold.  
That's why we limited our consideration by pump amplitude below $g=2\cdot C_{s} $, since with a further increase in the amplitude, there is a rapid increase in the average photon number, but quantum correlations are becoming less and less pronounced.\par
Finally we would like to point out the differences between degenerate~($\omega = 0$) and general cases.
On Fig.~\ref{plot:coherent} different columns correspond to these different cases.
The first thing that catches your eye when analyzing the presented distributions is that in general case there is a pronounced entanglement between the signal and idle modes in the central waveguide.
The value of logarithmic negativity for these modes is several times higher than the value for the nearest optical fibers with indexes 1 and -1.
Moreover, the stationary value of logarithmic negativity monotonically increases with the increase of pump amplitude~(Fig.~\ref{plot:coherent}(f) green dashed line).
If we compare the distributions for the general and degenerate cases excluding the diagonal elements from consideration, we will see a qualitative similarity.
All scales, as well as the type of dependencies on the parameters of the problem are the same.
However, there are quantitative differences.
Again, we emphasize that these differences are connected not with detuning from exact synchronism, but with differences in the modes structure for the degenerate and general cases.
\subsection{Entanglement evolution: phase noise influence}
In this subsection, we consider the influence of phase noise in the pump on the entangled states formation in the array of optical waveguides.
First of all, we note that we will consider extremely narrow-band noise.
The spectral width of the pump is several orders of magnitude smaller than the parametric resonance width.
This noise has no significant effect on the evolution of the average photon number.
Therefore, all the results from the first subsection remain valid in the discussed case of partially coherent pump.
In this case, as well as in paper~\cite{martynov_influence_2017}, entanglement turns out to be very sensitive to phase fluctuations. 
Also, as was shown above, the differences between the degenerate and general cases are small, so in this section we will discuss results only for the degenerate case. \par
  
\begin{figure}[t]
	\begin{minipage}[t]{1.0\linewidth}
		\includegraphics[scale=1.0]{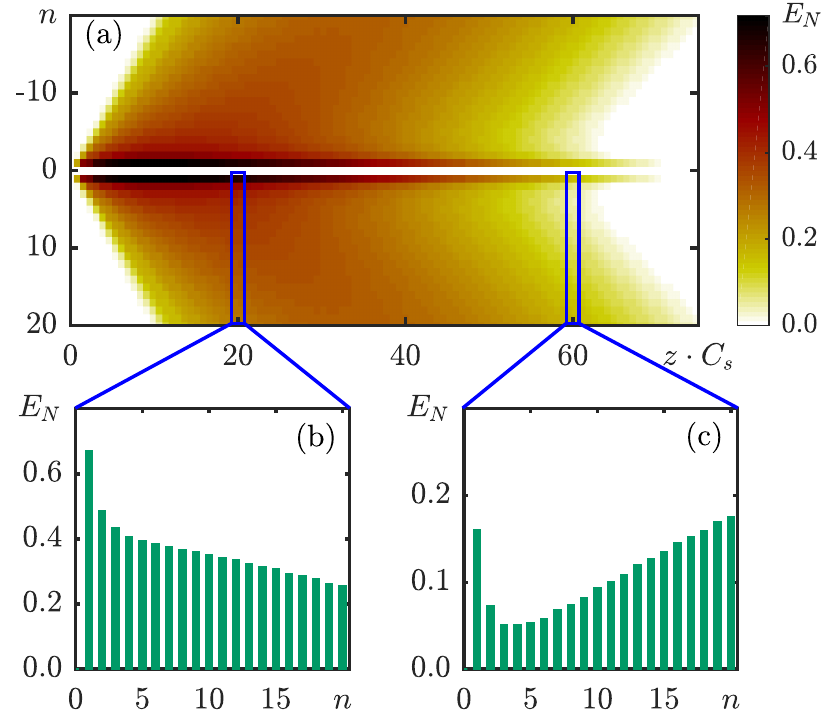}\vspace{-1.25mm}
		\caption{(a) Evolution of logarithmic negativity $E_N$, calculated for optical modes of waveguides with indexes $n$ and $-n$ in case when phase noise is present in the pump. 
		On (b) and (c) distributions of logarithmic negativity, formed at the distance (b) $z\cdot C_{s} = 20$ and (c) $z\cdot C_{s} = 60$ from the input of an array, are presented.
		Calculations are performed for parameters $g=C_{s}$, $\Delta\omega = 10^{-4}\cdot C_{s}\cdot v_{p}$.}
		\label{plot:noise:1}
	\end{minipage}
\end{figure}
\begin{figure}[t]
	\begin{minipage}[t]{1.0\linewidth}
		\includegraphics[scale=1.0]{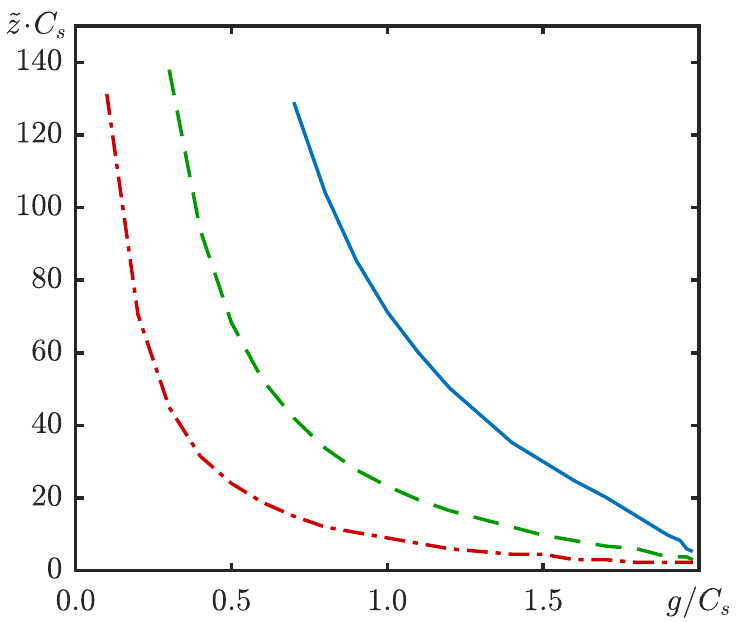}\vspace{-1.25mm}
		\caption{Dependency on the pump amplitude $g$ of the distance $\tilde{z}$, for which weveguides with indexes $1$ and $-1$ are entangled. 
		Results are presented for three values of pump spectral width: $\Delta\omega = 10^{-4}\cdot C_{s}\cdot v_{p}$ - blue solid line, $\Delta\omega = 10^{-3}\cdot C_{s}\cdot v_{p}$ - green dashed, $\Delta\omega = 10^{-2}\cdot C_{s}\cdot v_{p}$ - red dash-dotted.}
		\label{plot:noise:2}
	\end{minipage}
\end{figure}
As it was shown in the previous section, in the discussed system, quantum correlations (entanglement) are most pronounced for optical waveguides located symmetrically with respect to the central one.
On Fig.~\ref{plot:noise:1}(a) logarithmic negativity evolution for waveguides with indexes $n$ and $-n$ is shown.  
One may notice that for small distances phase fluctuations do not have a noticeable effect.
Distribution on Fig.~\ref{plot:noise:1}(b) fully corresponds to the distribution in Fig.~\ref{plot:coherent}(b).
At greater distances, the value of logarithmic negativity start to decrease~(Fig.~\ref{plot:noise:1}(c)). 
On certain length it turns to zero, and it means that entanglement in the system disappear.
Thus we can determine distance $\tilde{z}$, for which $E_N$ differs from zero.
An interesting fact is that this quantity $\tilde{z}$ has the minimum value for pairs of waveguides with indexes $3..5$~(indexes slightly differs for different parameters of the system).\par
Dependency of entanglement existence distance on pump amplitude is shown on Fig.~\ref{plot:noise:2}.
One can notice that $\tilde{z}$ decrease rapidly with pump amplitude increase.
When the threshold value is crossed, the distance practically vanishes.
After crossing threshold value $g=2\cdot C_{s}$, the distance practically vanishes.
We believe that this is due to a sharp increase in the growth rate of the average number of photons. The more photons are in fibers, the more destructive impact phase fluctuations have on the entanglement in the array.
If, for example, we consider the experiment described in paper~\cite{solntsev_generation_2014}, for parameters of their setup $C_{s}\approx 1cm^{-1}$, $g\sim 10^{-3} \cdot C_{s}$.
For this parameters if we assume the pump to have 1MHz spectral width, then $\tilde{z}$ will be approximately equal to several meters.
This value is many times greater than the length of the used lattice in the mentioned experiment.
At the same time, if we decide to obtain multiphoton entangled states and increase the pump amplitude to correspond to $g\approx 2\cdot C_{s}$, the length of the entanglement existence will decrease to an order of few centimeters.
This length corresponds to the length of the arrays of nonlinear waveguides currently used .
\section{Conclusion}
In this paper we have studied the process of entangled states formation as a result of spontaneous parametric down-conversion of light during propagation along a one-dimensional array of coupled nonlinear optical waveguides.
Such systems are currently being actively studied in the context of designing integrated optical circuits for quantum information processing based on quantum states of light.
The obtained results are important for the development of systems based on multimode squeezed states for quantum information processing.
In conclusion, we present the key results.\par
First, the evolution of the quantum state of light in an array is qualitatively different for different values of the parametric pump amplitude.
When a certain threshold defined by the geometry of the considered system is exceeded, the growth of the average number of photons changes from linear to exponential.
This above threshold generation regime is also characterized by less pronounced quantum correlations and is more susceptible to the influence of phase fluctuations present in parametric pump.
Thus, we focused on the parameter region corresponding to the pump amplitude below the threshold.\par
To study the evolution of entanglement present in the system, the distributions of the logarithmic negativity calculated for different pairs of optical waveguides were investigated.
At small propagation distances, the results obtained coincide with those given in~\cite{solntsev_generation_2014}. 
In this paper generation of entangled biphoton states of light in a similar system was studied, and for each waveguide it was shown the presence of entanglement with a large number of other waveguides.  
However, at large distances, when the states become substantially multi-photon, our calculations show that the entanglement remains only between pairs that are located symmetrically with respect to the pumped waveguide.
Moreover, the amount of entanglement gradually reaches a stationary value, although the average number of photons continues to grow.
Also there is an optimal ratio between the interaction coefficient and the pump amplitude, for which the stationary value of logarithmic negativity takes the maximum value.\par
Presence of phase noise in the pump qualitatively change the evolution of entanglement in the system.
At the beginning, as in the coherent case, the value of the logarithmic negativity grows and goes to the stationary value.
After that the value begins to decrease until it becomes zero, which indicates the complete disappearance of entanglement in the system.
Even the slightest phase noise leads to a finite distance at which quantum correlations are present in the system.
The specified distance essentially depends on the pump amplitude.
In the case of weak pumping, the distance of the existence of entanglement becomes large, in the limit infinite.
At the same time, as the amplitude approaches the threshold value, this distance becomes comparable to the characteristic interaction distance between the waveguides.
This circumstance indicates that the discussed effect does not influence significantly the generation of low-photon entangled states as in~\cite{solntsev_generation_2014,antonosyan_effect_2014, yang_manipulation_2014}.
But it may be the decisive limiting factor for generating entangled states, which are multimode squeezed.
\begin{acknowledgments}
This work was supported by the Ministry of Education and
Science of the Russian Federation under contract No.14.W03.31.0032.\par
\end{acknowledgments}
\section{Appendix. Reduction to oscillator interacting with non-Markovian reservoir.}
In this appendix we will reduce the equations~(\ref{equation:signal:4}) to a single one describing parametric oscillator interacting with non-Markovian reservoir.
First of all after substitution $\hat{a}^{s}_{n}\rightarrow\hat{a}^{s}_{n}e^{-i\omega/v_{s}}$:
\begin{equation}\label{eq:appendix:1}
\dfrac{\partial}{\partial z}\hat{a}^{s}_{n}\left(\omega\right) = iC_{s}\left(\hat{a}^{s}_{n-1}\left(\omega\right) +\hat{a}^{s}_{n+1}\left(\omega\right)\right) + ig\delta_{0n}e^{i\varphi\left(- z/v_{p}\right)}\hat{a}^{s\dag}_{n}\left(-\omega\right).
\end{equation}
Now we will introduce new operators:
\begin{equation}\label{eq:3:ancillary:2}
\begin{array}{cc}
\hat{d}_{n} = \hat{a}_{n}^{s}\left(\omega\right) + \hat{a}_{-n}^{s}\left(\omega\right) &,\;n>0,\\
\hat{d}_{n} = 0 &,\;n=0,\\
\hat{d}_{n} = -\hat{d}_{-n}&,\;n<0.
\end{array}
\end{equation} 
Evolution of this new operators is described by equations:
\begin{equation}\label{eq:3:evolution:3}
\left. \hat{d}_n \right. ^{'}_{z} =iC_s\left(\hat{d}_{n+1} + \hat{d}_{n-1}\right) + \hat{f}_n(z)\;,\; n\in \mathbb{Z}, 
\end{equation}
where 
\begin{equation}
\hat{f}_n(z) = \left\{\begin{array}{cc}
2iC_s\hat{a}^{s}_{0}\left(\omega\right) &,\;n=1,\\
-2iC_s\hat{a}^{s}_{0}\left(\omega\right) &,\;n=-1,\\
0 &,\;n\ne1,-1.
\end{array} \right.
\end{equation} 
To solve equations~(\ref{eq:3:evolution:3}) we will build generating function $\hat{F} = \sum \hat{d}_{n}q^{n}$. This function satisfies the equation:
\begin{equation}\label{eq:generating:1}
\dfrac{\partial\hat{F}}{\partial z} = iC_{s}q'\hat{F} + 2iC_{s}\hat{a}^{s}_{0}\left(q - \dfrac{1}{q}\right),
\end{equation}
where $q' = q+\dfrac{1}{q}$. Solution of~(\ref{eq:generating:1}) has the next form:
\begin{equation}\label{sol:generating:1}
    \hat{F} = \hat{F}\left(0\right)e^{iC_{s}q'z} + 2iC_{s}\int_{0}^{z}\hat{a}^{s}_{0}\left(z'\right)\left(q-\dfrac{1}{q}\right)e^{iC_{s}q'\left(z-z'\right)}dz',
\end{equation}
Now we will use the generating function for Bessel functions:
\begin{equation}
e^{iC_{s}\left(q + \dfrac{1}{q} \right)z}=\sum_{k}J_k\left(2C_{s}z\right)\left(-iq\right)^{k}.
\end{equation}
Using this equality we can present first term in~(\ref{sol:generating:1}) as:
\begin{equation}
\sum_{m}\sum_{n}\hat{d}_m\left(0\right)J_{n-m}\left(2C_{s}z\right)\left(-i\right)^{n-m}q^{n},
\end{equation}
and the second term will take the form:
\begin{equation}
2iC_{s}\int_{0}^{z}\hat{a}^{s}_{0}\left(z'\right)\sum_{k}\left(-i\right)^{k}J_{k}\left(2C_{s}(z-z')\right)\left(q^{k+1} - q^{k-1}\right)dz'.
\end{equation}
Hence, the entire solution for $\hat{d}_{1}$ will be:
\begin{equation}
\hat{d}_{1} = -i\sum_m \hat{d}_{m}\left(0\right)J_{1-m}\left(2C_{s}z\right)\left(-i\right)^{-m} + 2iC_{s}\int_{0}^{z}\hat{a}^{s}_{0}\left(z'\right)\left(J_{0}\left(2C_{s}(z-z')\right) + J_{2}\left(2C_{s}(z-z')\right)\right)dz'.
\end{equation}
After substitution of this solution into equation for $\hat{a}^{s}_{0}$ from~(\ref{eq:appendix:1}) we will get the closed equation for light field in central waveguide:
\begin{equation}\label{appendix:result}
\dfrac{\partial}{\partial z}\hat{a}^{s}_{0}\left(\omega\right) = -2C_{s}\int_0^{z}K\left(z-z'\right)\hat{a}^{s}_{0}\left(z'\right)dz' +ige^{i\varphi\left(- z/v_{p}\right)}\hat{a}^{s\dag}_{0}\left(-\omega\right)+\hat{\mathcal{F}}\left(z\right),
\end{equation}
where 
\begin{equation}
K\!\!\left(x\right)\!=\!\! J_{0}\left(2C_{s}x\right)+J_{2}\left(2C_{s}x\right)
\end{equation}
and
\begin{equation}
\begin{array}{c}
\hat{\mathcal{F}}\!\left(z\right)\!=\!-C_{s}\sum_{p=1}^{\infty}\left(\hat{a}_{p}^{s}(0,0)\!+\!\hat{a}_{-p}^{s}(0,0)\!\right)\\
\left(\!e^{-ip\pi/2}\!J_{1-p}\!\left(2C_s z\right)\!-\! e^{ip\pi/2}\!J_{1+p}\!\left(2C_s z\right)\!\right).
\end{array}
\end{equation}
Equation~(\ref{appendix:result}) looks similar to the one describing damped parametric oscillator~\cite{scully_quantum_1997}. 
The only difference is that obtained equation interaction occur with a non-Markovian reservoir.
The first term in~(\ref{appendix:result}) is a relaxation operator. 
And $\hat{\mathcal{F}}\left(z\right)$ is an Langevin source, which is required to preserve commutation relations~\cite{scully_quantum_1997, tokman_purcell_2019}.
\bibliographystyle{apsrev4-1}
\bibliography{FibersParametric}
\end{document}